\documentclass[global,twocolumn]{revtex4}

\usepackage{graphics}
\usepackage{epstopdf}
\usepackage{amsmath}
\usepackage{amssymb}
\usepackage{gensymb}
\usepackage{amsmath}
\usepackage{soul}
\usepackage{hyphenat}

\begin{document}
\title{Improved measurement of the hyperfine structure of the laser cooling level $4f^{12}(^3 H_6)5d_{5/2}6s^2$ $(J=9/2)$ in $^{169}$Tm}
\author{S.A.Fedorov, G.A.Vishnyakova, E.S.Kalganova, D.D.Sukachev, A.A.Golovizin, D.O.Tregubov, K.Yu.Khabarova, A.V.Akimov, N.N.Kolachevsky, V.N.Sorokin
}                     
\affiliation{P.N.Lebedev Physical Institute of the Russian Academy of Sciences, Leninsky prosp. 53, 119991 Moscow, Russia, Moscow Institute of Physics and Technology (State University), Instituskiy per. 9, 141700 Dolgoprudny, Moscow Region, Russia, and Russian Quantum Center, Business-center ''Ural'', ul. Novaya 100A, 143025, Skolkovo, Moscow region, Russia}

\keywords{Thulium atom, hyperfine structure, frequency modulation saturated absorption spectroscopy}

\begin{abstract}
We report on the improved measurement of the hyperfine structure of $4f^{12}(^3 H_6)5d_{5/2}6s^2$ $(J=9/2)$ excited state in Tm-169 which is involved in the second-stage laser cooling of Tm. To measure the absolute value of the hyperfine splitting interval we used Doppler-free frequency modulation saturated absorption spectroscopy of Tm atoms in a vapor cell. The sign of the hyperfine constant was determined independently by spectroscopy of laser cooled Tm atoms. The hyperfine constant of the level equals $A_J=-422.112(32)$ MHz that corresponds to the energy difference between two hyperfine sublevels of $-2110.56(16)$~MHz. In relation to the saturated absorption measurement we quantitatively treat contributions of various mechanisms into the line broadening and shift. We consider power broadening in the case when Zeeman sublevels of atomic levels are taken into account. We also discuss the line broadening due to frequency modulation and relative intensities of transitions in saturated-absorption experiments. 
\end{abstract}
\maketitle

\section{Introduction}
\label{sec:Intro}
Along with some other lanthanides Tm possesses a large ground state magnetic moment of 4 Bohr magnetons, that makes ultracold Tm atoms an attractive object for study of dipolar interactions~\cite{TmCoolingPRA}. Besides that Tm has a narrow magnetic-dipole transition at 1.14~$\mathrm{\mu}$m which makes it a favorable candidate for optical clock applications~\cite{TmSecCoolLasPhys,TmClock}.\par

The nuclear spin of the only stable thulium isotope $^{169}$Tm equals $I=1/2$. Each level of the electronic structure is split into two hyperfine components with energy shifts described solely in terms of the magnetic dipole constant $A_J$~\cite{Sobelman}:
\begin{equation}
\label{eq:AJDef}
\Delta E_F=\frac{1}{2}h A_J\left(F(F+1)-I(I+1)-J(J+1)\right),
\end{equation}
where $\hbar$ is the reduced Planck constant, $F$ and $J$ are the total atom and electron moments, correspondingly.

The hyperfine (HF) structure of Tm atoms was extensively studied with a variety of techniques. The hyperfine splitting (HFS) of the ground level was previously determined with a precision of about 1~kHz using double-resonance spectroscopy~\cite{TmGNDHfs,TmMetastHfs}. Since the double-resonance method is applicable only to ground and metastable states, the HF structure of the excited states was mainly studied by interferometry~\cite{prevRes} and laser spectroscopy in a wide spectral range. Laser spectroscopy methods implemented in~\cite{TmHfsMeas2,TmHfsMeas3} provided an accuracy typically not exceeding a few megahertz.\par 

\begin{figure}
\begin{center}
\resizebox{0.9\columnwidth}{!}{
  \includegraphics{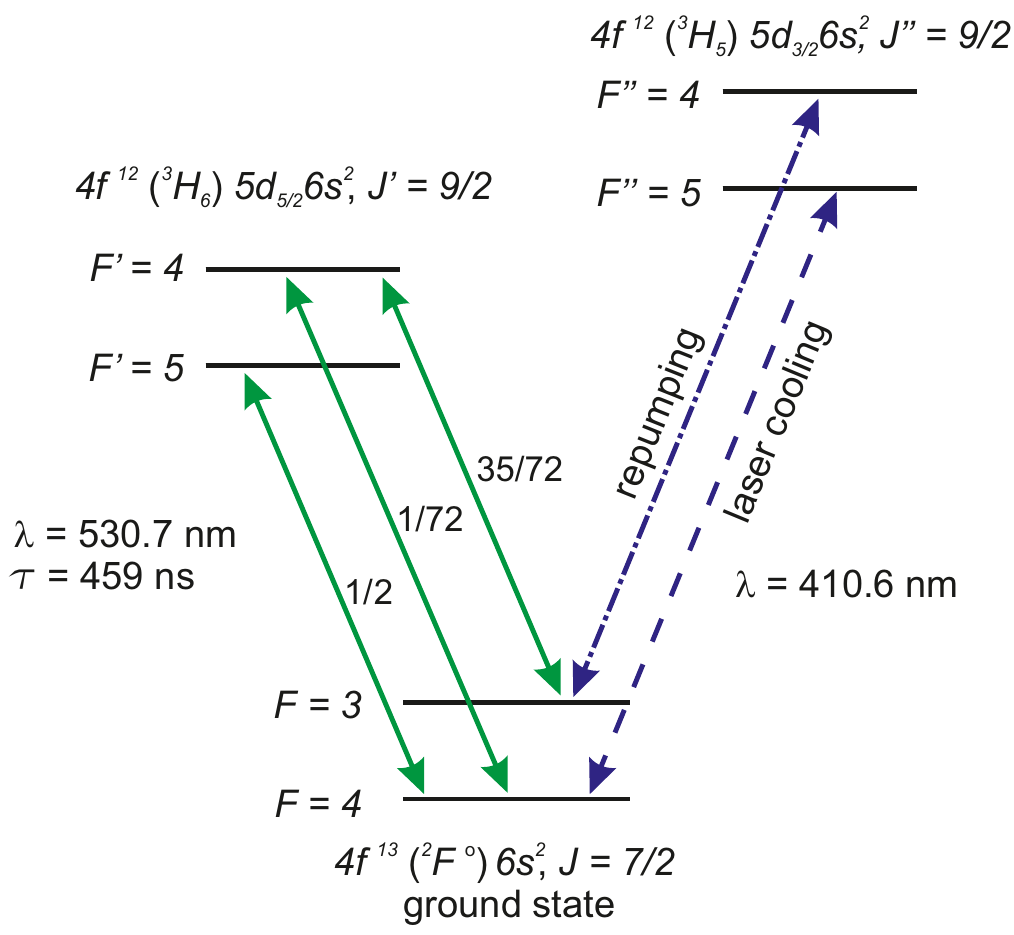}
}

\caption{The relevant transitions in $^{169}$Tm. The transitions at 530.7~nm are shown in solid (green) lines, the numbers indicate the relative transition strengths. The 410.6~nm transitions used for magneto-optical trapping and repumping (See sec.~\ref{sec:HfsOrder}) are shown in dashed (blue) lines.}
\end{center}
\label{fig:RelLev}       
\end{figure}

The level $4f^{12}(^3 H_6)5d_{5/2}6s^2$ $(J=9/2)$ is coupled to the Tm ground state via 530.7~nm transition (fig.~\ref{fig:RelLev}) with the natural line width of $\Gamma/2\pi=347(17)$~kHz~\cite{TmLifeTimes}. The transition between the ground state ($F=4$) and the investigated state ($F'=5$) is fully cyclic and is used for the second-stage laser cooling of Tm atoms down to the temperature of 10~$\mathrm{\mu}$K~\cite{TmSecCoolLasPhys}.\par

Here we report the hyperfine splitting measurement of this level within an improved uncertainty of 160~kHz using the frequency modulation saturated absorption spectroscopy. \par

The paper is organized as follows: sec.~\ref{sec:HfsOrder} describes how we determine the sign of the hyperfine constant, sec.~\ref{sec:HfsAbsSplitting} describes the measurement of the hyperfine constant, in sec.~\ref{sec:FittingAndResults} we present the line shape model and in sec.~\ref{sec:FMSatAbsLineBr} we describe the sources of uncertainty. In App.~\ref{sec:Disc} optical pumping effects are discussed.

\section{Spectroscopy in the Tm magneto-optical trap}
\label{sec:HfsOrder}

\begin{figure*}
	\begin{center}
	\resizebox{0.95\textwidth}{!}{
		\includegraphics{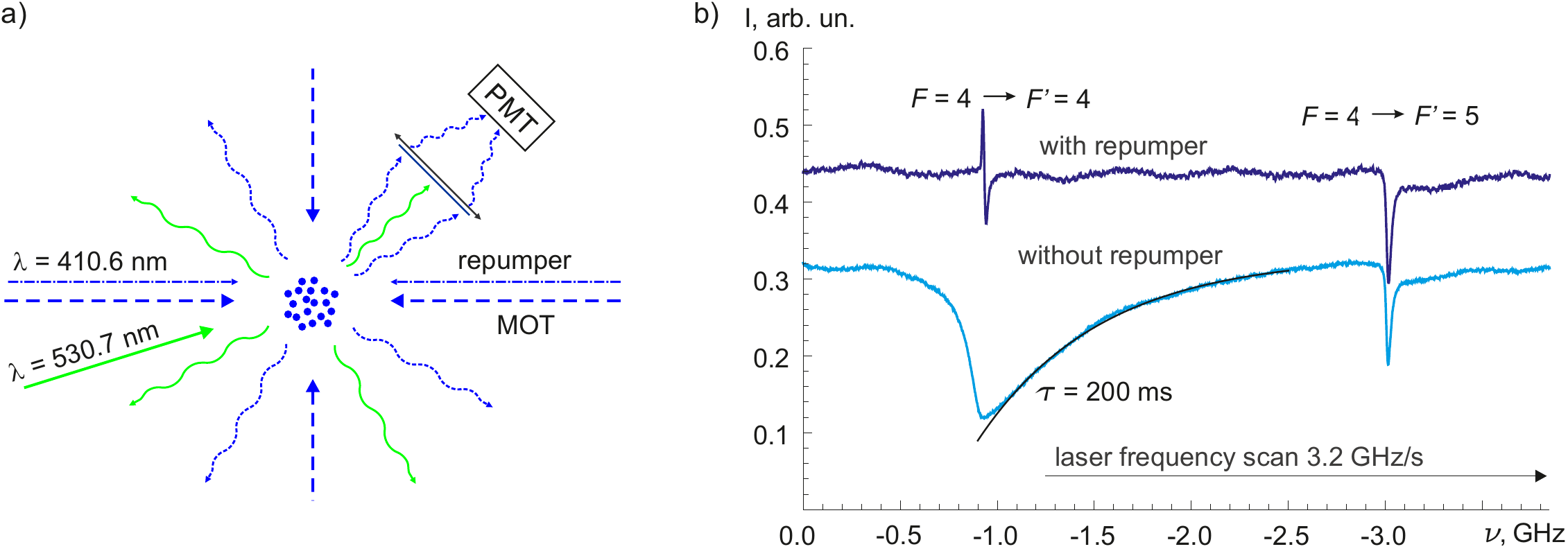}
	}
	\caption{(a) Laser cooled Tm atoms is irradiated with 530.7~nm laser. PMT is photomultiplier tube. (b) Fluorescence intensity at the wavelength 410.6~nm of the atoms in an operating "blue" magneto-optical trap being irradiated by 30~mW of the 530.7~nm laser beam. The frequency on the 530.7 nm was continuously scanned as indicated at the figure, the frequency scale was calibrated using an auxiliary Fabry-Perot cavity. The lower and upper curves correspond to the measurements without and with repumping laser (fig.~\ref{fig:RelLev}). The solid black curve is the exponential fit to the data.}
	\end{center}
	\label{fig:MOTfluorescence}       
\end{figure*}

In linear spectroscopy the transitions between different hyperfine components may be readily identified according to their relative intensities described in terms of $6j$-symbols. The situation is more complicated in saturated absorption spectroscopy, where considerable optical pumping occurs between the hyperfine components. The relative intensities become sensitive to the experimental conditions (see App.~\ref{sec:Disc} for discussion). Taking into account the importance of the hyperfine components order for laser cooling applications we determined it by an independent method.\par

Tm atoms were laser cooled and trapped in a ''blue'' magneto-optical trap (MOT) operating at 410.6~nm almost cyclic transition $F=4\to F''=5$ between ground and $4f^{12}(^3 H_5)\allowbreak 5d_{3/2}6s^2$ $(J=9/2)$ levels (See fig.~\ref{fig:RelLev})~\cite{TmCoolingPRA}. We irradiated the atoms trapped in the operating ''blue'' MOT with 30~mW (10~W/cm$^2$) of 530.7~nm laser (second harmonic of Toptica DL-pro) and recorded the fluorescence signal at 410.6~nm by photomultiplier tube (See fig.~\ref{fig:MOTfluorescence}(a)). The fluorescence at steady-state was proportional to the number of atoms in the MOT. In contrast to alkali atoms, Tm MOT at 410.6~nm normally can operate without a repumping laser~\cite{TmCoolingPRA} because the cooling radiation acts as a repumper itself and return atoms which were optically pumped to $F=3$ back to $F=4$ via $F''=4$.\par

 Frequency scan of 530.7 nm laser across $F=4\to F'=4$ and $F=4\to F'=5$ resonances resulted in corresponding dips in the MOT fluorescence (See fig.~\ref{fig:MOTfluorescence}(b), lower curve). We identified two transitions by the different fluorescence dynamics as shown in fig.~\ref{fig:MOTfluorescence}(b). Excitation of an atom to the $F'=4$ component of the upper level would predominantly be followed by a spontaneous decay to $F=3$ sublevel of the ground state (fig.~\ref{fig:RelLev}) and loss of the atom from the MOT. After the  laser frequency was tuned out of this resonance, the population of MOT recovered with the trap loading time constant (200~ms). In contrast, when the laser frequency was scanned through the cyclic $F=4\to F'=5$ transition, the number of atoms in the MOT remained unchanged and the fluorescence signal recovered promptly. The latter happened because exited atoms were slow enough to stay in the trapping region (the radiative decay time $1/\Gamma$ of investigated level is about 460 ns). \par

To verify our understanding of the underlying processes we repeated the described experiment, but applied an additional repumping laser beam of 2.4~mW (1~W/cm$^2$) at 410.6~nm that pumped the atoms from the hyperfine $F=3$ sublevel back to $F=4$ of the ground state through the $F''=4$ sublevel (See fig.~\ref{fig:RelLev}). As expected, the repumping laser prevented the loss of atoms from MOT when 530.7~nm laser was tuned to $F=4\to F'=4$ resonance and the MOT fluorescence recovered equally fast for both observed resonances (See fig.~\ref{fig:MOTfluorescence}(b), upper curve). \par

We conclude that the sublevel $F'=5$ lies lower than the $F'=4$ one. This fact allows to implement second-stage laser cooling without repumping laser~\cite{TmCoolingPRA}.

\section{Saturated absorption experiment}
\label{sec:HfsAbsSplitting}         

\begin{figure}
\begin{center}
\resizebox{0.9\columnwidth}{!}{
  \includegraphics{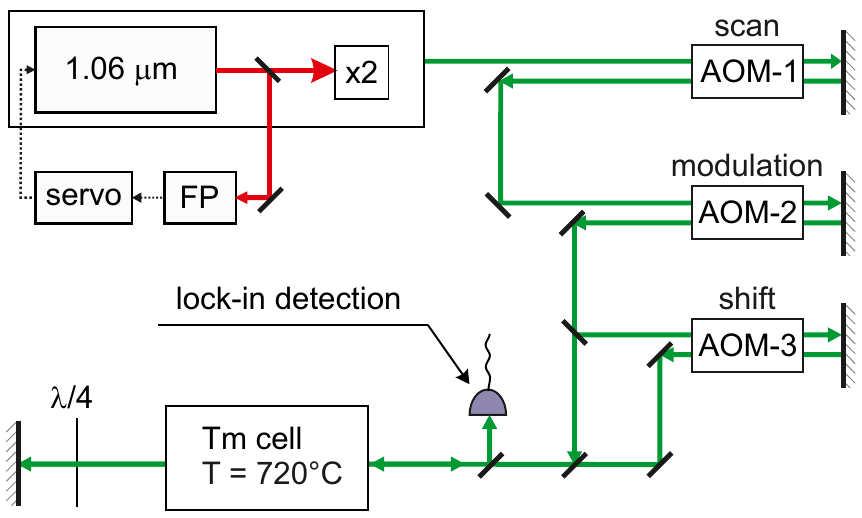}
}
\caption{Schematics of saturated absorption experiment. Acousto-optical modulator (AOM-1) was used for the frequency scan, AOM-2 for the modulation and AOM-3 for the frequency shift of one laser beam to bridge the gap between $F=4\to F'=5$ and $F=3\to F'=4$ transitions. All AOMs are working in double-pass configurations. The laser is locked to a stable high-finesse Fabri-Perot cavity (FP).}
\end{center}
\label{fig:SAsetup}       
\end{figure}

The described above MOT-loss measurement of HFS did not provide enough accuracy due to the complicated resonance profiles and insufficiently accurate frequency scan calibration. For more accurate measurement we used frequency modulation saturated absorption spectroscopy in a vapor cell~\cite{SatAbs}. \par

The experimental setup is shown in fig.~\ref{fig:SAsetup}. Tm vapor stainless steel cell at the temperature of about 720\degree~C was enclosed in a magnetic shield. To stabilize the frequency of the 530.7~nm laser we locked it to a high-finesse Fabry-Perot cavity using Pound-Driver-Hall technique~\cite{PDH}. Stabilization provided the expected laser line width below 100~Hz~\cite{PoundDreverHall}. Long-term drift of Fabri-Perot cavity was measured to be 1.3~kHz/s by monitoring the frequency interval between the $F=4\to F'=5$ resonance and a particular TEM$_{00}$ mode of the cavity.\par      

In our configuration the probe beam was formed by reflecting the pump beam. Since saturated absorption is accompanied by power broadening of the transition, it was advantageous to reduce the laser power to the lowest level. To detect weak absorption signals we used frequency modulation of the laser beam and lock-in detection. The modulation at the frequency of 60 kHz was applied using the acousto-optical modulator AOM-2 (See fig.~\ref{fig:SAsetup}). Central frequency of AOM-2 was 203~MHz. \par

Both $F=3$ and $F=4$ hyperfine components of the ground state were initially populated almost equally at the cell temperature of 720\degree~C. We extracted the frequency difference of the transitions $F=4\to F'=5$ and $F=3\to F'=4$ from the lock-in amplified saturated absorption signal.\par

\begin{figure*}
\resizebox{\textwidth}{!}{
  \includegraphics{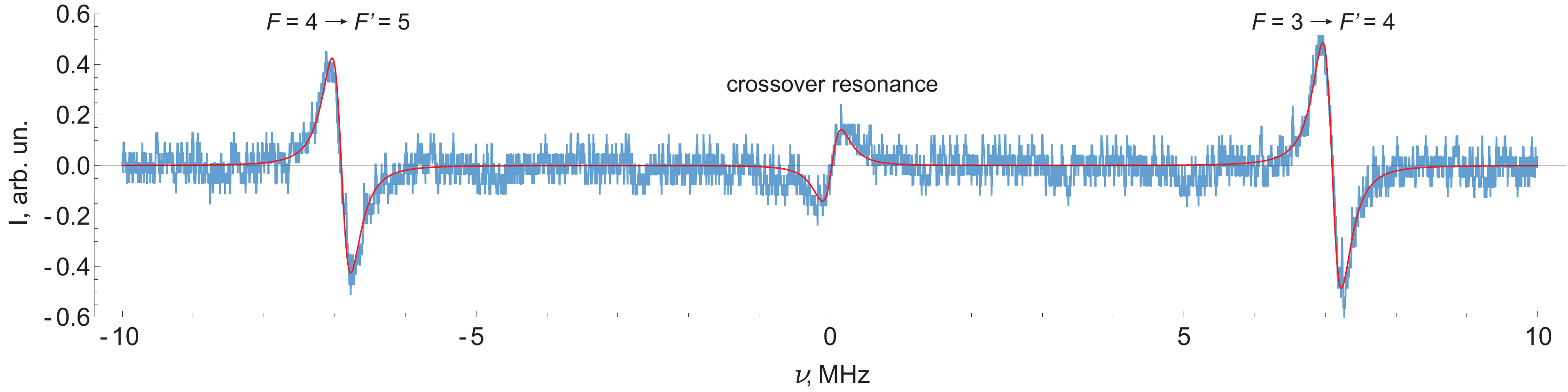}
}
\caption{Demodulated signal as a function of laser frequency detuning set by AOM-1. The experimental data with subtracted linear Doppler offset and the fit with a sum of three dispersion profiles with independent positions and amplitudes are shown. The scan rate was 0.5~MHz/s.}
\label{fig:scan}  
\end{figure*}

For frequency scanning we used AOM-1 driven by signal of RF-generator centered at 211~MHz. Scan range of AOM was about 50~MHz and did not allow to cover the 614~MHz frequency gap between the $F=4\to F'=5$ and $F=3\to F'=4$ transitions. To observe both transitions simultaneously, we used bichromatic radiation with one of the beams being frequency-shifted by an auxiliary AOM-3 for 600~MHz (See fig.~\ref{fig:SAsetup}). A typical scan for both $F=4\to F'=5$ and $F=3\to F'=4$ transitions with 14~MHz separation is shown in fig.~\ref{fig:scan}. At the picture the left resonance is formed by unshifted laser beam and the right is by frequency-shifted one. The crossover resonance at the picture is a feature of the saturated absorption method that results from the interaction of atoms with both beams~\cite{SatAbs}.  \par

Both laser beams had powers of $100(10)$~$\mathrm{\mu W}$ and the $1/e^2$ intensity radii of $\approx0.55$~cm. This corresponds to saturation parameter $I/I_s=0.35$, where 

\begin{equation}
I_s=\pi h c\Gamma/3\lambda^3=304\,\mathrm{\mu W/cm^2}
\end{equation}
 is the transition saturation intensity. The extinction coefficient was about 30\% for both beams.\par

The hyperfine constant $A_J$ was deduced from the experimental frequency splitting between $F=4\to F'=5$ and $F=3\to F'=4$ transitions and the known value of the ground state hyperfine splitting of 1496.550(1)~MHz~\cite{TmGNDHfs}. The line shape model used for data analysis is described in the next sec.~\ref{sec:FittingAndResults}.

\section{The line shape model} 
\label{sec:FittingAndResults}

Frequency modulation in nonlinear spectroscopy may result in complicated line shapes of the observed transitions. A number of analytical results describing line shapes in some specific experimental conditions are presented in~\cite{OptHetSatSp,HetSatAbsPumpFMMod,CohEffFMSpec}. In the current setup we used low modulation frequency $\omega_m< \Gamma$ and small modulation index $M\approx 0.1 \ll 1$, that allows to describe the observed signals in terms of a steady-state absorption function $\alpha(\omega)$~\cite{Myers1959}. If the phase of the laser beam is modulated
\begin{equation}
\label{eq:laserPhase}
\phi=M\sin(\omega_m t),
\end{equation} 
the absorption will adiabatically follow the instantaneous laser frequency, producing the signal $S_{\omega_m}$
\begin{equation}
\label{eq:AdFMResp}
S_{\omega_m}=M\omega_m \frac{d \alpha(\omega)}{d\omega}\sin(\omega_m t).
\end{equation}         

The regime of low frequency modulation allows for the straightforward interpretation of the observed signals in a wide range of experimental conditions. In particular, the eq.~\ref{eq:AdFMResp} imposes no limitations to the laser power, total energy absorption or the details of nonlinear interaction. Moreover, the various mechanisms of line broadening are naturally taken into account by eq.~\ref{eq:AdFMResp}, in contrast to more sophisticated models of Refs.~\cite{OptHetSatSp,HetSatAbsPumpFMMod,CohEffFMSpec,ShBrWaveFrCurvTheor}. \par

To describe the experimental signal we chose $\alpha(\omega)$ to be the Lorenzian function with the same width for all three resonances (See fig.~\ref{fig:scan}). This choice is justified by the fact that the observed spectral linewidths are only about 30\% larger than their natural linewidths.\par

\section{Line shifts and broadening}
\label{sec:FMSatAbsLineBr}

A number of effects may lead to shift and broadening of resonances in satureted-absorption spectorscopy. These effects may introduce error in measurements and have to be carefully analyzed. Below we discuss the shift and broadening mechanisms that took place in our experiment and estimate associated uncertainties in HFS measurement.\par

\subsection{Power broadening}

In saturated absorption spectroscopy the detected signal rapidly decreases as the pump and probe intensities go below the saturation intensity $I_s$. Operation at high laser intensities, however, results in power broadening of the observed resonances. In the case of atomic transitions between levels with angular moments $F$ and $F'$ the power-broadened linewidth $\Gamma_{\mathrm{p.b.}}$ may be calculated as
\begin{equation}
\Gamma_{\mathrm{p.b.}}=\Gamma\sqrt{1+a(F,F')I/I_s}.
\end{equation} 
Here $I$ is the laser beam intensity and $a$ is a numerical factor. In a well-known case of two-level system $a=1$. However, as showed our numerical calculations based on direct Bloch equations solution, for cyclic atomic transitions $F'=F+1$ with $F,F'\ge 2$ in linearly polarized light $a$ approaches 0.5. \par

The origin of the reducing factor $a$ is best illustrated in the case of atom interacting with a linearly polarized laser beam and the quantization axis being directed along the light polarization~\cite{HypLevPumping2}. For the multilevel atom every upper magnetic sublevel is populated only by absorption from the ground sublevel with the same magnetic number, $m$. Absorption rate $\gamma_{|F,m\rangle\to|F',m\rangle}$ is reduced compared to these of two-level atom by the corresponding Clebsch-Gordan coefficient $C$
\begin{equation}
\gamma_{|F,m\rangle\to|F',m\rangle}=\frac{s}{2}(C^{F',m}_{F,m;1,0})^2\Gamma\left(\mathrm{\Pi}(g_m)-\mathrm{\Pi}(e_m)\right).
\end{equation} 
Here $s$ is the saturation parameter
\begin{equation}
s=I/I_s\frac{(\Gamma/2)^2}{\Delta\omega^2+(\Gamma/2)^2},
\end{equation} 
and $\mathrm{\Pi}(g_m)$ and $\mathrm{\Pi}(e_m)$ are the populations of the lower and upper magnetic components. If we consider a particular case of $F=0\to F=1$ transition, we will find the only relevant coefficient $C^{1,0}_{0,0;1,0}=1$ and will reproduce the result for a two-level system $\Gamma_{\mathrm{p.b}}=\Gamma\sqrt{1+I/I_s}$ with $a$ being 1. However for transitions between levels with higher $F$ the situation is different. In this case $(C^{F',m}_{F,m;1,0})^2<1$ and larger light intensity is needed in order to make the stimulated transition rates comparable with the spontaneous decay rates. \par  

In the case of our experiment the power broadened linewidth is calculated to be 380~kHz at $I/I_s= 0.35$. We note, that intensity here is the single-beam intensity, since at moderate powers the width of the saturated-absorption resonances is the average of the power-broadened widths corresponding to the pump and the probe intensities~\cite{ShBrWaveFrCurv}.\par
  
\begin{table}
\caption{Line broadening effects for 530.7~nm transitions.}
\label{tab:LineBrContrib}      
\begin{tabular}{ll}
\hline\noalign{\smallskip}
Source & Contribution, kHz\\
\noalign{\smallskip}\hline\noalign{\smallskip}
Natural linewidth & 347(17)\\
Power broadening & 30\\
Wavefront curvature & 150\\
Beams misalignment & 100\\
Frequency modulation  & 20\\
Time-of-flight broadening & 7\\
Zeeman splitting & $<10$\\
Collisional broadening & $<40$\\
Photon recoil & 8\\
\noalign{\smallskip}\hline
Observed linewidth & 450(50)~kHz\\
\noalign{\smallskip}\hline
\end{tabular}
\end{table}

\subsection{Time-of-flight broadening}

The transition time for atoms passing through the laser beam in our experiment was 50 times larger compared to the radiative lifetimes of the upper states. This resulted in additional broadening  of 7~kHz. 

\subsection{Effect of wavefront curvature and misalignment of beams}

Deviation of the pump and probe beams from counter-propagating plane waves generally results in shift and broadening of saturated-absorption resonances. The finite absolute curvatures of the wavefronts and slight misalignment of beams may lead to residual Doppler broadening (so called "geometrical broadening"). In particular, the pump and probe beams misalignment (b.m.) of finite angle $\beta$ leads to the Gaussian broadening of~\cite{ChebotayevLetokhov}
\begin{equation}
\delta \omega_{\mathrm{b.m.}} \simeq \beta k u_{D}\text{, }u_{D}=\left(\frac{2 k_B T}{M}\right)^{1/2}.
\end{equation}
Here $k$ is the light wavevector and $u_{D}$ is the Doppler velocity of atoms. For our setup $u_D=310$~m/s and the beams misalignment was estimated to be $< 2\cdot 10^{-4}$~rad that corresponds to the broadening of about $\delta \omega_{\mathrm{b.m.}}/2\pi=100$~kHz. \par

In our experiment the time of flight of atoms through the laser beams is much longer than the inversed natural linewidth. In this case  the broadening caused by the wavefront curvature (w.c.) of the Gaussian beam also has essentially the same geometrical origin and may be estimated as~\cite{ChebotayevLetokhov}
\begin{equation}
\label{eq:geomBroaden}
\delta \omega_{\mathrm{w.c.}} \simeq \frac{ku_D}{\sqrt{k z_R}}, 
\end{equation}    
where $z_R$ is the Rayleigh length, assumed to be the same for pump and probe beams. For the reported measurement $z_R$ was slightly different for the two perpendicular crossections of the beams with average values of the order of 1~m. The corresponding broadening $\delta \omega_{\mathrm{w.c.}}/2\pi=150$~kHz.\par

In the absence of reflection symmetry between the pump and probe beams shifts of the saturated absorption resonances take place ~\cite{ShBrWaveFrCurvTheor}. We conservatively estimate the line shifts in our measurement to be of the same value as the broadening in eq.\ref{eq:geomBroaden}.

\subsection{Broadening due to frequency modulation}

As was mentioned in the sec.~\ref{sec:FittingAndResults}, we used frequency modulation with the ratio $\omega_m/\Gamma\approx0.2$. Frequency modulation may cause additional effects compared to the adiabatic picture described by eq.~\ref{eq:AdFMResp}. In this section we compare the predictions of eq.~\ref{eq:AdFMResp} with more rigorous results given in~\cite{OptHetSatSp}. 

Frequency modulation results in a line shape, that often may be described as some combination of power broadened  Lorenzian and dispersion profiles~\cite{OptHetSatSp,LinFMSpectrTheor}.
\begin{equation}
L(\omega)=\frac{(\Gamma_{\mathrm{p.b.}}/2)^2}{\Delta\omega^2+(\Gamma_{\mathrm{p.b.}}/2)^2} \text{, } D(\omega)=\frac{\Delta\omega\Gamma_{\mathrm{p.b.}}/2}{\Delta\omega^2+(\Gamma_{\mathrm{p.b.}}/2)^2},
\end{equation}  
where $\Delta\omega=(\omega-\omega_0)$ is the laser detuning from the atomic resonance. In the saturated absorption configuration the in-phase signal is predicted to be~\cite{OptHetSatSp}
\begin{equation}
\label{eq:rSatAbsLinePh}
S_{\mathrm{ph.}}\propto\left[L(\omega+\omega_m/2)-L(\omega-\omega_m/2)\right]\sin(\omega_m t).
\end{equation} 

If $\omega_m\lesssim\Gamma$, the eq.~\ref{eq:rSatAbsLinePh} predicts signals almost indistinguishable from $dL(\omega)/d\omega$, but with a slightly larger effective line width. For our conditions we found the line broadening to be about 10~kHz.  \par   

We note, that at finite modulation frequency also an in-quadrature signal presents
\begin{equation}
\label{eq:rSatAbsLineQ}
S_{\mathrm{q.}}=\left[D(\omega-\omega_m/2)-2D(\omega)+D(\omega-\omega_m/2)\right]\cos(\omega_m t).
\end{equation}
The small admixture of quadrature component results in a slight variation in the detected line width and line shape with the demodulation signal phase change. For the present experiment, however, the effect of quadrature component is of the same negligible order ($\sim10$~kHz) as the correction to the in-phase component by eq.~\ref{eq:rSatAbsLinePh}.\par
 Resonance broadening by modulation is not accompanied by shift, and thus introduce error only via statistical uncertainty in determining the peak positions.

\subsection{Magnetic fields}

In order to reduce the influence of magnetic field the vapor cell was enclosed into a magnetic shield. In addition, the Zeeman splitting was greatly suppressed by similar Lande $g$-factors of the lower and the upper transition levels ($\Delta g\approx 0.01$). Conservative estimate of the Zeeman shift is less then 10~kHz.\par

\subsection{Collisional and photon recoil effects}

We made an upper estimate of the collisional effects by measuring the positions of atomic resonances at different vapor temperatures 640--730\degree~C (corresponding vapor pressure range $6\times10^{-5}$--$8\times10^{-4}$~mbar). No collisional broadening was observed at the level   of 100 MHz/mbar and no collisional shift was detected with uncertainty 50 MHz/mbar. Thus, the contribution of collisional effects into the line width was below 40 kHz and the contribution into the HFS measurement error was below 20 kHz.\par 
The photon recoil effect \cite{RadPressEff} was $\hbar k^2/2\pi M=8$~kHz.

\begin{table}
\caption{Errors budget for the HFS measurement.}
\label{tab:errBudget}       
\begin{tabular}{ll}
\hline\noalign{\smallskip}
Source & Uncertainty, kHz\\
\noalign{\smallskip}\hline\noalign{\smallskip}
Statistics & 20\\
Wavefront curvature & 150\\
Zeeman splitting & $10$\\
Collisional shift & $20$\\
ULE cavity drift & $30$\\
\noalign{\smallskip}\hline
Total uncertainty & 160~kHz\\
\noalign{\smallskip}\hline
\end{tabular}
\end{table}

\subsection{Uncertainty and results}
 The results for the line broadening and the HFS measurement uncertainty are summarized in  tab.~\ref{tab:LineBrContrib} and tab.~\ref{tab:errBudget} correspondingly.\par

The natural linewidth of the transition was consistently reported in a few works, the most recent result being $\Gamma/2\pi=347(17)$~kHz~\cite{TmLifeTimes}. From our experimental data fit we extracted the observed line width of $450(50)$~kHz, which provides a reasonable agreement between the estimated and observed broadened lines.\par

We estimated the error in the HFS measurement as root sum square of the contributions listed in the tab.~\ref{tab:errBudget}. It can be seen that the net uncertainty of 160~kHz was dominated by the shifts of resonances due to wavefront curvature, while the statistical error did not exceed 20~kHz. Another source of uncertainty was a long-term drift of Fabri-Perot cavity used for laser frequency stabilization. Typical drift rate was 1.3~kHz/s yielding 30~kHz shift for the time of measurement.\par 

An independent method to assess the measurement error was provided by the crossover resonance position. Comparison of the measured crossover resonance frequency and its expected value relative to the transitions peaks $(\nu_{F=3\to F'=4}-\nu_{F=4\to F'=5})/2$ resulted in the discrepancy of 100~kHz, which is consistent with the error estimate given in the tab.~\ref{tab:errBudget}.\par

We measured the measured frequency difference between the investigated transitions to be 
\begin{equation}
\nu_{F=3\to F'=4}-\nu_{F=4\to F'=5} = 614.01(16)~\mathrm{MHz.}
\end{equation} 
Energy difference between $F'=5$ and $F'=4$ hyperfine sublevels is
\begin{equation}
E_{F'=5}-E_{F'=4}=-2110.56(16)~\mathrm{MHz.}
\end{equation} 
Finally, the improved value for hyperfine constant $A_J$ of the level $4f^{12}(^3 H_6)5d_{5/2}6s^2$ ($J=9/2$) equals (See eq.~\ref{eq:AJDef})
\begin{equation}
A_J=-422.112(32)~\mathrm{MHz.}
\end{equation} 

The latter result is consistent with the value, previously reported by Kuhl~\cite{prevRes}, but is 20 times more accurate.

\section{Conclusions} 
We reported the measurement of the hyperfine structure of $4f^{12}(^3 H_6)5d_{5/2}6s^2$ $(J=9/2)$ level in Thulium atom, which is used for second-stage laser cooling. \par 

The hyperfine splitting of the level was previously measured within uncertainty of 3~MHz using interferometery in atomic beam~\cite{prevRes}. The result of our measurement agrees with the previous result, but the accuracy is improved more than an order in magnitude. The reported in~\cite{prevRes} value of HFS constant $A_J$ agreed with theoretical calculations presented in~\cite{hfsTheorCalc} within the measurement uncertainty, and thus our new value may provide a better error estimate for theoretical calculations. \par
 
The dominant source of uncertainty in or measurement was the frequency shift of saturated-absorption resonances due to the wavefront curvature. We also estimated contributions of other line shift and broadening mechanisms, in particular, line broadening due to frequency-modulation and power broadening in the case of transition between degenerate atomic levels. The total estimated broadening agrees with the experimental observation. \par

\section{Acknowledgments}
This work was supported by RFBR grants 15-02-05324a, 15-02-03936a and the program of fundamental researches of the RAS "Extreme light fields and its applications".

\appendix
\section{Effect of optical pumping on intensity relation of transitions between hyperfine sublevels}
\label{sec:Disc}

In linear spectroscopy the intensities of transitions between hyperfine components of two levels are given by their line strengths $S$
\begin{equation}
\label{LineStrength6j}
S\propto(2F_d+1)(2F_u+1)
\begin{Bmatrix}
 J_u & F_u& I\\
F_d& J_d& 1 \\
\end{Bmatrix}^2, 
\end{equation}
where $u$ and $d$ denotes the upper and lower levels of the transition. Using this relation the transitions between different hyperfine components may be readily identified in experiment. However, in saturated-absorption experiments the relation~(\ref{LineStrength6j}) may yield qualitatively wrong results. \par

Our experiment provides an example when the observed most intense line is not the one expected from eq.~\ref{LineStrength6j}.We calculate the transition $F'=5\to F=4$ to have the largest line strength 
\begin{equation}
\left(\frac{I_{F'=5\to F=4}}{I_{F'=4\to F=3}}\right)_{\mathrm{theor}}=\frac{44}{35}>1.
\end{equation}
In contrast, from the experiment we find (see fig.~\ref{fig:scan})
\begin{equation}
\left(\frac{I_{F'=5\to F=4}}{I_{F'=4\to F=3}}\right)_{\mathrm{exp}}\approx0.9<1.
\end{equation}

The observed intensities appear to be the consequence of optical pumping (for more discussion of the effect see Refs.~\cite{HypLevPumping,HypLevPumping2}). When the non-cyclic $F=3\to F'=4$ resonance is exited some fraction of atoms spontaneously decay from the upper $F'=4$ to the lower $F=4$ sublevel. In this way the number of atoms interacting with light effectively decreases, thus giving additional contribution to the Lamb dip and increasing the resonance amplitude. For the cyclic $F=4\to F'=5$ resonance the pumping is absent and its amplitude is of purely saturational origin. \par  

The order of magnitude estimation of the enhancement factor $f$ for the $F'=4\to F=3$ transition intensity is given by the ratio of atoms pumped into $F=4$ state during the time of flight:
\begin{equation}
f=1+\Gamma_{F'=4\to F=4}\times\tau_{\mathrm{flight}}.
\end{equation} 
This estimation gives a factor of 2.5, that is reasonably agrees with the experimental value of 1.5.\par

\end{document}